\documentclass[%
 reprint,
 superscriptaddress,
 amsmath,
 amssymb,
 aps,
 pra,
 floats,
 floatfix
]{revtex4-1}

\usepackage{graphicx} 
\usepackage[breaklinks=true]{hyperref} 
\usepackage{physics}
\usepackage{xcolor}
\usepackage{bbold}
\usepackage{mathtools}
\usepackage{siunitx}
\usepackage{microtype}
\usepackage{comment}
\usepackage{algorithm}
\usepackage{algpseudocode}
\usepackage[capitalize]{cleveref}

\DeclareSIUnit{\au}{{a.u.}}
\usepackage{comment}

\newcommand{\HF}{\text{HF}}

\newcommand{\EOM}{\,\text{EOM}\!}
\newcommand{\LR}{\,\text{LR}\!}

\begin{document}

\title{Simulating weak-field attosecond processes 
with a Lanczos reduced basis approach to time-dependent equation-of-motion coupled-cluster theory}

\author{Andreas S. \surname{Skeidsvoll}}
\thanks{%
These authors contributed equally
}%
\affiliation{%
Department of Chemistry, Norwegian University of Science and Technology, NO-7491 Trondheim, Norway
}%
\author{Torsha \surname{Moitra}}
\thanks{%
These authors contributed equally
}%
\affiliation{%
DTU Chemistry—Department of Chemistry, Technical University of Denmark, DK-2800 Kongens Lyngby, Denmark
}%

\author{Alice \surname{Balbi}}
\affiliation{%
Scuola Normale Superiore, Piazza dei Cavalieri, 7, IT-56126 Pisa, PI, Italy
}

\author{Alexander C. \surname{Paul}}
\affiliation{%
Department of Chemistry, Norwegian University of Science and Technology, NO-7491 Trondheim, Norway
}

\author{Sonia \surname{Coriani}}
\email{Electronic mail: soco@kemi.dtu.dk}
\affiliation{%
DTU Chemistry—Department of Chemistry, Technical University of Denmark, DK-2800 Kongens Lyngby, Denmark
}
\affiliation{%
Department of Chemistry, Norwegian University of Science and Technology, NO-7491 Trondheim, Norway
}

\author{Henrik \surname{Koch}}
\email{Electronic mail: henrik.koch@sns.it}
\affiliation{%
Department of Chemistry, Norwegian University of Science and Technology, NO-7491 Trondheim, Norway
}
\affiliation{%
Scuola Normale Superiore, Piazza dei Cavalieri, 7, IT-56126 Pisa, PI, Italy
}

\date{\today}

\begin{abstract}
A time-dependent equation-of-motion coupled-cluster singles and doubles (TD-EOM-CCSD) method is implemented, which uses a reduced basis calculated with the asymmetric band Lanczos algorithm. The approach is used to study weak-field processes in small molecules induced by ultrashort valence pump and core probe pulses. We assess the reliability of the procedure by comparing TD-EOM-CCSD absorption spectra to spectra obtained from the time-dependent coupled-cluster singles and doubles (TDCCSD) method, and observe that spectral features can be reproduced for several molecules, at much lower computational times. We discuss how multiphoton absorption and symmetry can be handled in the method, and general features of the core-valence separation (CVS) projection technique. We also model the transient absorption of an attosecond X-ray probe pulse by the glycine molecule.

\end{abstract}

\maketitle

\section{Introduction}
Stimulated by the recent experimental realization of various laser pulses with durations on the attosecond (\SI{1e-18}{\second}) time scale~\cite{Hassan_2016_Nature_attosec_exp,Galli_2019_subfs_exp,Fabris_2015_NaturePhoton_attosec_exp,Duris2020,PhysRevResearch.2.042018}, capable of monitoring electronic motion, the theoretical simulation of coherent electron dynamics is currently an active field of research~\cite{Li_2020_CHemRev_RTmethods}.

Real-time electronic structure theory considers the explicit time dependence of the electronic system by evolving the time-dependent Schr\"{o}dinger equation in the time domain~\cite{Li_2020_CHemRev_RTmethods}. Explicitly time dependent methods can directly provide the time-domain evolution of electronic wave functions together with nuclear motion, representing a versatile way of tracking real-time ultrafast dynamics and phenomena in both perturbative and non-perturbative regimes.~\cite{doi:10.1021/acs.chemrev.7b00423,doi:10.1021/acs.jctc.0c00730}

The development of real-time methods commenced in the late 70s and early 80s in the field of nuclear physics~\cite{TDHF-Koonin,3D-TDHF-Bonche,3D-TDHF-Flocard}. Despite these early endeavours, real-time methods did not become practical at that time due to the lack of electron correlation effects at the Hartree-Fock level and the high computational cost associated with propagation of correlated wave functions. However, decades of steady advancements in computing power and numerical algorithms have led to a renewed interest in explicit time propagation in correlated methods like density functional theory~\cite{https://doi.org/10.1002/qua.25096,Lian_2018_RTTDDFT}, multiconfigurational self-consistent-field~\cite{TD-CASSCF-PhysRevA.87.062511,TD-CASSCF-PhysRevA.88.023402,TD-CASSCF-PhysRevA.89.063416}, configuration interaction~\cite{TD-CI-10.1063/1.1999636,TD-CI-doi:10.1021/jp107384p,TD-CI-doi:10.1080/00268976.2012.675448,TD-CI-PhysRevLett.107.196806}, algebraic diagrammatic construction~\cite{CEDERBAUM1999205,D0FD00104J} and coupled-cluster~\cite{Eugene_2017_JPCL_NEXAFS_TDEOMCC,Koulias_2019_JCTC_relativstic_rtTDEOMCC,QD-CC-doi:10.1063/1.4718427,TD-CC-10.1063/1.3530807,TD-CC-PhysRevC.86.014308,PhysRevA.102.023115,Park:2019,Pedersen:2019,TD-CC-10.1063/1.5142276,doi:10.1021/acs.jctc.0c00977,Park:2021}.

In this work, we present an implementation and representative case studies of the TD-EOM-CC model for simulating weak-field attosecond valence pump---core probe processes. In conjunction with a reduced-space band Lanczos algorithm for obtaining the valence and core excited states, this model offers results similar to its TDCC counterpart in weak fields, at significantly lower computational costs. The reduction in cost enables the study of larger systems. 

The article is organized as follows. In \cref{sec:theory} we detail the theory behind TD-EOM-CC and the asymmetric band Lanczos algorithm. Here, we also discuss a strategy used in order to guide the reduced space solver to directly obtain the transitions between excited states.
The computational procedure used is detailed in \cref{sec:computational}.
In \cref{sec:results}, simulations for various molecular systems are presented. First a benchmark study is presented for LiF, validating our proposed method. Second, the applicability of the core-valence separation scheme is tested for LiH. Then, two-photon absorption phenomenon has been captured using a stepwise procedure emulating the actual physical process for C$_2$H$_4$. Finally, we have put forward a theoretical assessment of pump-probe absorption for the glycine molecule, which is deemed suitable for further experimental investigations. The findings are summarized in \cref{sec:conclusion}.

\section{Theory}\label{sec:theory}
\subsection{System}\label{sec:system}
We model the system, comprising a molecule and its interaction with laser pulses, with the Hamiltonian
\begin{equation}
    H(t) = H^{(0)} + V(t)~,
\end{equation}
where the field-free Hamiltonian $H^{(0)}$ describes the molecule with fixed nuclei and without interactions with the external electromagnetic field. The semi-classical time-dependent interaction term, written in the dipole approximation and length gauge, is
\begin{equation}
    V(t) = - \vb*{d} \cdot \vb*{\mathcal{E}}(t)~,
\end{equation}
and describes the interaction between the molecular electrons and the external electromagnetic field. The latter is represented by the electric field $\vb*{\mathcal{E}}(t) = [\mathcal{E}_x(t) \; \mathcal{E}_y(t) \; \mathcal{E}_z(t)]^T$ and electronic dipole operator vectors, $\vb*{d} = [d_x \; d_y \; d_z]^T$. We assume that the molecule is initially in the ground state of the field-free Hamiltonian, and take the electric field to be a linear combination of the electric fields of any number of laser pulses,
\begin{equation}
    \vb*{\mathcal{E}}(t) = \sum_n \vb*{\mathcal{E}}_{0n}\cos(\omega_{0n}(t-t_{0n}) + \phi_n)f_n(t)~.
\end{equation}
The field of laser pulse $n$ has an associated carrier frequency $\omega_{0n}$, peak strength $\abs{\vb*{\mathcal{E}}_{0n}}$ and polarization $\vb*{\mathcal{E}}_{0n}/\abs{\vb*{\mathcal{E}}_{0n}}$, and an $8\sigma_n$-truncated Gaussian envelope function
\begin{equation}
    f_n(t) =
    \begin{cases}
    e^{-(t-t_{0n})^2/(2\sigma_n^2)}~\qc & \abs{t-t_{0n}} \leq 8\sigma_n~, \\
    0 & \text{otherwise}~,
    \end{cases}
\end{equation}
with duration specified by $\sigma_n$, the temporal root-mean-square (RMS) width. It is also specified by the central time $t_{0n}$ and the carrier-envelope phase (CEP) $\phi_n$. We assume the carrier-envelope phase to be zero for all pulses, meaning that the maximum values of the envelope and cosine carrier functions belonging to pulse $n$ coincide at $t_{0n}$.

The energy absorbed during the interaction can be given by~\cite{Wu_2016,PhysRevA.102.023115}
\begin{equation}
    \Delta E = \int_0^\infty \omega S(\omega)\dd{\omega}~,
\end{equation}
where $S(\omega)$ is the response function
\begin{equation}
    S(\omega) = -2\Im[\vb*{\ev*{\widetilde{d}}}\!(\omega)\cdot\vb*{\widetilde{\mathcal{E}}}{}^*(\omega)]~\qc \omega > 0~.
\end{equation}
The vectors $\vb*{\ev*{\widetilde{d}}}\!(\omega)$ and $\vb*{\widetilde{\mathcal{E}}}(\omega)$ are the Fourier transforms of the time-dependent dipole moment expectation value and electric field vectors, respectively, and the asterisk denotes complex conjugation. A positive or negative value of the function $S(\omega)$ describes the probability of absorption or emission of light with frequency $\omega$, respectively~\cite{Wu_2016}.

\subsection{TD-EOM-CC states}\label{sec:td-eom-cc}
The time-dependent ket and bra of a TD-EOM-CC state can be expressed as
\begin{gather}
    \ket{\Psi(t)} = \sum_i \ket{\psi_i}s_i(t)~\qc \bra*{\widetilde{\Psi}(t)} = \sum_i k_i(t)\bra*{\widetilde{\psi}_i}~,
\end{gather}
where the italic indices $i,j$ are used to denote general EOM-CC states, including the ground state with index 0. The time-independent EOM-CC kets and bras are given by
\begin{gather}
    \label{eq:eomstates}
    \ket{\psi_i} = e^T R_i\ket{\HF}~\qc \bra*{\widetilde{\psi}_i} = \bra{\HF}L_ie^{-T}~.
\end{gather}
We assume that the EOM-CC states are biorthonormal
\begin{equation}
    \label{eq:statebiorthogonality}
    \bra*{\widetilde{\psi}_i}\ket{\psi_j} = \delta_{ij}~.
\end{equation}

In the following, we let the indices $\kappa, \lambda$ denote general determinants in the projection space, including the reference Hartree-Fock determinant with index 0. We use the indices $\mu, \nu$, on the other hand, to denote excited determinants.

The cluster operator $T$ and the right and left operators $R_i, L_i$ of \cref{eq:eomstates} can be expressed as linear expansions in a finite set of operators $\tau_\kappa, \tau_\kappa^\dagger$,
\begin{gather}
    \label{eq:operators}
    T = \sum_\mu \tau_\mu t_\mu~\qc R_i = \sum_\kappa\tau_\kappa r_{\kappa i}~\qc L_i = \sum_\kappa l_{i \kappa} \tau_\kappa^\dagger~,
\end{gather}
where the operator with index 0 is the unit operator,
\begin{equation}
    \tau_0 = \tau_0^\dagger = 1~,
\end{equation}
and the $\tau_\mu$ and $\tau_\mu^\dagger$ operators generate excited determinants from the ket and bra reference Hartree-Fock determinants, respectively,
\begin{gather}
    \tau_\mu\ket{\HF} = \ket{\mu}~\qc \bra{\HF}\tau_\mu^\dagger = \bra{\mu}~, \\
    \tau_\mu^\dagger\ket{\HF} = 0~\qc \bra{\HF}\tau_\mu = 0~.
\end{gather}
We assume that the determinants are biorthogonal,
\begin{equation}
    \bra{\kappa}\ket{\lambda} = \delta_{\kappa\lambda}~.
\end{equation}

If all possible electronic excitations are included in the summations in \cref{eq:operators}, the method is equivalent to full configuration interaction (FCI). The sum can also be restricted to given excitation levels, giving approximate methods that scale polynomially with the system size. This includes the coupled-cluster singles and doubles (CCSD) method, where summation is only done over single and double excitations. We do not explicitly state the excitation levels included in following expressions, since they hold for both restricted and unrestricted summation.

The cluster amplitudes $t_\mu$ in \cref{eq:operators} can be found from solving equations involving the similarity-transformed field-free Hamiltonian operator $\bar{H}^{(0)}$ projected onto the right reference and left excited determinants
\begin{equation}
    \bra{\mu}\bar{H}^{(0)}\ket{\HF} = 0~,
\end{equation}
where the similarity transformation of an operator $X$ is denoted by an overbar,
\begin{equation}
    \bar{X} = e^{-T}Xe^{T}~.
\end{equation}

After the optimal cluster amplitudes $t_\mu$ have been determined, the right and left vectors of EOM-CC state $i$, with components $r_{\kappa i}$ and $l_{i\kappa}$, can be found as right and left eigenvectors of the field-free Hamiltonian matrix, with elements
\begin{equation}
    \label{eq:hamiltonianmatrix}
    H_{\kappa\lambda}^{(0)} = \bra{\kappa}\bar{H}^{(0)}\ket{\lambda}~.
\end{equation}

The right and left eigenvectors of the matrix in \cref{eq:hamiltonianmatrix} with the lowest eigenvalue, specifying the ground EOM-CC state with index $0$, have the following structure
\begin{gather}
    \label{eq:rightgscomponents}
    r_{00} = 1~\qc r_{\mu0} = 0~, \\
    \label{eq:leftgscomponents}
    l_{00} = 1~\qc l_{0\mu} = \bar{t}_\mu~.
\end{gather}
The multipliers $\bar{t}_\mu$ are solutions to the equations
\begin{equation}
    \bra{\HF}\bar{H}^{(0)}\ket{\nu} + \sum_\mu\bar{t}_\mu A_{\mu\nu}^{(0)} = 0~,
\end{equation}
where elements of the field-free coupled-cluster Jacobian matrix $\vb*{A}^{(0)}$ are given by
\begin{equation}
    A_{\mu\nu}^{(0)} = \bra{\mu}\comm*{\bar{H}^{(0)}}{\tau_\nu}\ket{\HF}~.
\end{equation}
The other right and left eigenvectors of the matrix in \cref{eq:hamiltonianmatrix} correspond to excited EOM-CC states, denoted by the italic indices $m,n$. The eigenvectors have the following reference determinant components
\begin{gather}
    \label{eq:rightrefcomponent}
    r_{0m} = -\sum_\mu\bar{t}_\mu r_{\mu m}~, \\
    \label{eq:leftrefcomponent}
    l_{m0} = 0~.
\end{gather}
These components enforce the biorthogonality between the ground and excited states, in accordance with \cref{eq:statebiorthogonality}. The vectors $\vb*{R}_m$ and $\vb*{L}_m$, containing the components $r_{\mu m}$ and $l_{m\mu}$ of excited EOM-CC state $m$, are right and left eigenvectors of $\vb*{A}^{(0)}$.

\subsection{Derivation of TD-EOM-CC equations}
The time derivative of the coefficients of the TD-EOM-CC ket can be found from projecting the ket time-dependent Schr\"{o}dinger equation (TDSE)
\begin{equation}
    \iota\pdv{t}\ket{\Psi(t)} = H(t)\ket{\Psi(t)}~,
\end{equation}
where $\iota$ denotes the imaginary unit, onto the bra of EOM-CC state $i$, giving
\begin{equation}
    \label{eq:righttdeomcc}
    \iota\pdv{s_i(t)}{t} = \sum_j H_{ij}(t)s_j(t)~,
\end{equation}
where the matrix elements of an operator $X(t)$ are given by
\begin{equation}
    \label{eq:matrixelements}
    X_{ij}(t) = \bra*{\widetilde{\psi}_i}X(t)\ket{\psi_j}~.
\end{equation}
Likewise, the time derivative of the coefficients of the TD-EOM-CC bra can be found from projecting the bra TDSE,
\begin{equation}
    -\iota\pdv{t}\bra*{\widetilde{\Psi}(t)} = \bra*{\widetilde{\Psi}(t)}H(t)
\end{equation}
onto the ket of EOM-CC state $j$, giving
\begin{equation}
    \label{eq:lefttdeomcc}
    -\iota\pdv{k_j(t)}{t} = \sum_i k_i(t) H_{ij}(t)~.
\end{equation}

To our knowledge, \cref{eq:righttdeomcc} and \cref{eq:lefttdeomcc} were first presented in Ref.~\cite{TD-CI-doi:10.1021/jp107384p}, and were also used in Ref.~\cite{TD-CI-doi:10.1080/00268976.2012.675448}. After time-dependent coefficients have been obtained from solution of these equations, the time-dependent expectation value of an operator $X(t)$ can be calculated according to
\begin{equation}
    \ev{X}\!(t) = \sum_{ij} k_i(t)X_{ij}(t)s_j(t)~.
\end{equation}

\subsection{Asymmetric band Lanczos}\label{sec:lanczos}
The asymmetric band Lanczos algorithm is a generalization of the simple asymmetric Lanczos algorithm, employing multiple starting vectors instead of single ones~\cite{doi:10.1137/1.9780898719581.ch7,Ove-and-Ian-bandLanczos,freund_2003}. The asymmetric square matrix $\vb*{M}$ and the sequences $(\vb*{b}_1, \ldots, \vb*{b}_m)$ and $(\vb*{c}_1, \ldots, \vb*{c}_p)$ of right and left starting vectors, respectively, determine the following sequences of vectors
\begin{gather}
    \begin{split}
        &(\vb*{\beta}_i(\vb*{M}, \vb*{b}_1, \ldots, \vb*{b}_m))_{i=1}^j \\
        &= (\underbrace{\vb*{b}_1,, \ldots, \vb*{b}_m, \vb*{M}\vb*{b}_1, \ldots, \vb*{M}\vb*{b}_m, \vb*{M}^2\vb*{b}_1, \ldots}_j)~,
    \end{split} \\
    \begin{split}
        &(\vb*{\gamma}_i(\vb*{M}^T,\vb*{c}_1, \ldots, \vb*{c}_p))_{i=1}^j \\
        &= (\underbrace{\vb*{c}_1, \ldots, \vb*{c}_p, \vb*{M}^T\vb*{c}_1, \ldots, \vb*{M}^T\vb*{c}_p, (\vb*{M}^T)^2\vb*{c}_1, \ldots}_j)~,
    \end{split}
\end{gather}
which can be linearly dependent. We define the right and left band Krylov subspaces as
\begin{gather}
    \begin{split}
        &\mathcal{K}_j(\vb*{M}, \vb*{b}_1, \ldots, \vb*{b}_m) \\
        &= \operatorname{span}\big\{(\vb*{\beta}_i(\vb*{M}, \vb*{b}_1, \ldots, \vb*{b}_m))_{i=1}^j\big\}~,
    \end{split} \\
    \begin{split}
        &\mathcal{K}_j(\vb*{M}^T,\vb*{c}_1, \ldots, \vb*{c}_p) \\
        &= \operatorname{span}\big\{(\vb*{\gamma}_i(\vb*{M}^T,\vb*{c}_1, \ldots, \vb*{c}_p))_{i=1}^j\big\}~,
    \end{split}
\end{gather}
of dimensions $n\leq j$. Note that these subspaces can also be regarded as block Krylov subspaces~\cite{freund_2003} whenever $j$ is a multiple of the number of start vectors.

The purpose of the asymmetric band Lanczos algorithm is to iteratively generate $n$ linearly independent right and left Lanczos vectors, $(\vb*{v}_1, \ldots, \vb*{v}_n)$ and $(\vb*{w}_1, \ldots, \vb*{w}_n)$, forming bases for the $n$-dimensional Krylov subspaces $\mathcal{K}_j(\vb*{M}, \vb*{b}_1, \ldots, \vb*{b}_m)$ and $\mathcal{K}_j(\vb*{M}^T,\vb*{c}_1, \ldots, \vb*{c}_p)$, respectively.
The $n$ right and left Lanczos vectors can be generated with the recurrence relations~\cite{doi:10.1137/1.9780898719581.ch7}
\begin{gather}
    \label{eq:trecursion}
    \begin{split}
        &\vb*{M}\vb*{V}_n = \vb*{V}_n\vb*{T}_n + \vu*{V}_n^{\mathrm{C}} + \vu*{V}_n^{\mathrm{D}}~,
    \end{split} \\
    \label{eq:ttilderecursion}
    \begin{split}
        &\vb*{M}^T\vb*{W}_n = \vb*{W}_n\widetilde{\vb*{T}}_n + \vu*{W}_n^{\mathrm{C}} + \vu*{W}_n^{\mathrm{D}}~,
    \end{split}
\end{gather}
where the right and left Lanczos vectors form the matrices $\vb*{V}_n = \begin{bmatrix}\vb*{v}_1 & \cdots & \vb*{v}_n\end{bmatrix}$ and $\vb*{W}_n = \begin{bmatrix} \vb*{w}_1 & \cdots & \vb*{w}_n\end{bmatrix}$,
respectively.  The unnormalized vectors that form the nonzero columns of $\vu*{V}_n^{\mathrm{C}} = \begin{bmatrix} \vb*{0} & \cdots & \vb*{0} & \vu*{v}_{n+1} & \cdots & \vu*{v}_{n+m_c} \end{bmatrix}$ and $\vu*{W}_n^{\mathrm{C}} = \begin{bmatrix} \vb*{0} & \cdots & \vb*{0} & \vu*{w}_{n+1} & \cdots & \vu*{w}_{n+p_c}\end{bmatrix}$ serve as candidates for the next right and left Lanczos vectors, respectively. The sparse matrices $\vu*{V}_n^{\mathrm{D}}$ and $\vu*{W}_n^{\mathrm{D}}$ contain unnormalized candidates from previous iterations that have been deflated (i.e. rejected) due to linear dependence on already accepted right and left Lanczos vectors, respectively. In numerical implementations, vectors are usually deflated when linear independence is below a given threshold, since inexact arithmetic prevents the description of exact linear dependence. The numbers $m_c$ and $p_c$, initially equal to the number of right and left starting vectors $m$ and $p$, give the current number of vectors available for deflation. We say that the sequence $(\vb*{\beta}_i(\vb*{M}, \vb*{b}_1, \ldots, \vb*{b}_m))_{i=1}^j$ or $(\vb*{\gamma}_i(\vb*{M}^T,\vb*{c}_1, \ldots, \vb*{c}_p))_{i=1}^j$ is fully exhausted when $m$ or $p$ deflations have occurred, respectively. The iterative procedure is then stopped, giving equal numbers of left and right Lanczos vectors.

The non-zero elements of $\vb*{T}_n$ in \cref{eq:trecursion} and of $\widetilde{\vb*{T}}_n$ in \cref{eq:ttilderecursion}, as determined by Algorithm 5.1 of Ref.~\cite{freund_2003}, enforce the biorthogonality between the $m+p+1$ vectors that can overlap in exact arithmetic~\cite{doi:10.1137/1.9780898719581.ch7}. Our implementation is consistent with this algorithm, with the exception that the following procedure is added to the beginning of Step (1)
\begin{algorithmic}
\If{$n > 1$}
    \For{$k = 1$ \textbf{to} $\max\{1, n-p_c-1\}$} 
        \State $\vu*{v}_n \gets \vu*{v}_n - \vb*{v}_k\big(\vb*{w}_k^T\vu*{v}_n\big)$
    \EndFor
\EndIf
\end{algorithmic}
and the following to the beginning of Step (2)
\begin{algorithmic}
\If{$n > 1$}
    \For{$k = 1$ \textbf{to} $\max\{1, n-m_c-1\}$} 
        \State $\vu*{w}_n \gets \vu*{w}_n - \big(\vu*{w}_n^T\vb*{v}_k\big)\vb*{w}_k$
    \EndFor
\EndIf
\end{algorithmic}
These additions lead to an algorithm that enforces the biorthogonality between all Lanczos vectors in inexact arithmetic,
\begin{equation}
    \vb*{W}_n^T\vb*{V}_n = \vb*{\Delta}_n = \operatorname{diag}(\delta_1, \delta_2, \ldots, \delta_m)~,
\end{equation}
and not just between the vectors that can overlap in exact arithmetic. We observe that this modification of the algorithm is important for the numerical stability when the number of iterations becomes large, but the modification also makes the number of vector operations substantially higher. The number of operations can potentially be reduced in future implementations, e.g., by formulating a restarted asymmetric band Lanczos algorithm, based on existing approaches~\cite{SHIMIZU2019372}.

We stop the algorithm when it reaches a given chain length $n$ (i.e. maximum number of iterations). The algorithm then generates, as Algorithm 5.1 of Ref.~\cite{freund_2003}, the $n\times n$ matrices
\begin{gather}
    \begin{split}
        \vb*{T}_n^{\mathrm{P}} &= \vb*{\Delta}_n^{-1}\vb*{W}_n^T\vb*{M}\vb*{V}_n \\
        &= \vb*{T}_n + \vb*{\Delta}_n^{-1}\vb*{W}_n^T \vb*{V}_n^\mathrm{D}~,
    \end{split} \\
    \begin{split}
        \widetilde{\vb*{T}}_n^\mathrm{P} &= \big(\vb*{W}_n^T\vb*{M}\vb*{V}_n\vb*{\Delta}_n^{-1}\big)^T \\
        &= \widetilde{\vb*{T}}_n + \big(\big(\vb*{W}_n^D)^T\vb*{V}_n\vb*{\Delta}_n^{-1}\big)^T~,
    \end{split}
\end{gather}
related by
\begin{equation}
    \vb*{\Delta}_n \vb*{T}_n^\mathrm{P} = \big(\widetilde{\vb*{T}}_n^\mathrm{P}\big)^T \vb*{\Delta}_n~,
\end{equation}
which are banded when no deflations have occured~\cite{doi:10.1137/1.9780898719581.ch7}.

The matrix $\vb*{T}_n^\mathrm{P}$ can be viewed as the oblique projection of $\vb*{M}$ onto $\mathcal{K}_j(\vb*{M}, \vb*{b}_1, \ldots, \vb*{b}_m)$ and orthogonally to $\mathcal{K}_j(\vb*{M}^T,\vb*{c}_1, \ldots, \vb*{c}_p)$~\cite{freund_2003}. Diagonalization of the matrix yields $n$ eigenvalues, which approximate the eigenvalues of $\vb*{M}$, and associated right and left eigenvectors. The right eigenvectors can be transformed to approximate eigenvectors of $\vb*{M}$ by premultiplication by $\vb*{V}_n$, and the left eigenvectors to approximate eigenvectors of $\vb*{M}^T$ by premultiplication by $\vb*{W}_n^T\vb*{\Delta}_n^{-1}$~\cite{doi:10.1137/1.9780898719581.ch7}. Approximate eigenvectors with dominant (low- and high-lying) eigenvalues are typically better converged than the ones in the middle~\cite{CullumLanczos,Golub:MatComp,SaadSparseBook}.

\subsubsection{Choice of starting vectors}\label{sec:startingvectors}
In \cref{sec:matrixelements}, we demonstrate how the expressions for matrix elements between excited states $m$ and ground or excited initial states are linear in the excited determinant components $r_{\mu m}$ and $l_{m\mu}$. This observation has previously been used to construct starting vectors based on excited states in EOM-CC theory~\cite{bruno-transient}, and have therein been used to justify the use of the starting vectors that we employ in the band Lanczos algorithm~\cite{Ove-and-Ian-bandLanczos,Sonia-cc-lanczos,bruno-transient}.

Starting vectors based on the ground state are given by
\begin{gather}
    \label{eq:b0}
    b_{\mu0}^X = \xi_\mu^X~, \\
    \label{eq:c0}
    c_{0\nu}^X = {}^{\EOM}\eta_\nu^X - X_{00} \bar{t}_\nu~,
\end{gather}
and starting vectors based on excited state $m$ are given by~\cite{bruno-transient}
\begin{gather}
    \label{eq:bm}
    b_{\mu m}^{X} = \sum_\nu\Big({}^{\EOM}A_{\mu\nu}^X + \delta_{\mu\nu}X_{00} - \xi_\mu^X\bar{t}_\nu\Big)r_{\nu m}~, \\
    \label{eq:cm}
    c_{m\nu}^{X} = \sum_\mu l_{m\mu}\Big({}^{\EOM}A_{\mu\nu}^X + \delta_{\mu\nu}X_{00} - \xi_\mu^X\bar{t}_\nu\Big)~.
\end{gather}
The specification of $\xi_\mu^X$, ${}^{\EOM}\eta_\nu^X$, $X_{00}$ and ${}^{\EOM}A_{\mu\nu}^X$ is given in \cref{sec:matrixelements}.

The starting vectors can be expressed in the Lanczos basis by inserting the resolution of identity in terms of the Lanczos vectors,
\begin{gather}
    \label{eq:bexpansion}
    \begin{split}
        \vb*{b}_i^{X} &= \sum_j \vb*{v}_{j}\big(\vb*{w}_j^T\vb*{b}_i^X\big) \\
        &= \sum_{j=1}^m \vb*{v}_jb_{ji}^{X}~,
    \end{split} \\
    \label{eq:cexpansion}
    \begin{split}
        (\vb*{c}_i^X)^T &= \sum_{j} \Big(\big(\vb*{c}_i^X\big)^T\vb*{v}_{j}\Big) \vb*{w}_j^T \\
        &= \sum_{j=1}^p c_{ij}^X\vb*{w}_j^T~.
    \end{split}
\end{gather}
The sum in \cref{eq:bexpansion} is restricted since $\vb*{b}_i^X \in \operatorname{span}\{\vb*{v}_{1}, \ldots, \vb*{v}_m\}$ while each $(\vb*{w}_{m+1}, \ldots)$ is biorthogonal to all $(\vb*{v}_{1}, \ldots, \vb*{v}_m)$. Likewise, the sum in \cref{eq:cexpansion} is restricted since $\vb*{c}_i^X \in \operatorname{span}\{\vb*{w}_{1}, \ldots, \vb*{w}_p\}$ while each $(\vb*{v}_{p+1}, \ldots)$ is biorthogonal to all $(\vb*{w}_{1}, \ldots, \vb*{w}_p)$.

Thus, the transition moments to excited states can be obtained by contracting the starting vectors with the vectors of excited state $n$,
\begin{gather}
    \label{eq:nitransition}
    \begin{split}
        X_{ni} &= \vb*{L}_n^T \vb*{b}_i^X \\
        &= \sum_{j=1}^m \big(\vb*{L}_n^T \vb*{v}_{j}\big) b_{ji}^{X} \\
        &= \sum_{j=1}^m L_{nj} b_{ji}^{X}~,
    \end{split} \\
    \label{eq:intransition}
    \begin{split}
        X_{in} &= \big(\vb*{c}_i^X\big)^T \vb*{R}_n \\
        &= \sum_{j=1}^p c_{ij}^X\big(\vb*{w}_j^T \vb*{R}_n\big) \\
        &= \sum_{j=1}^p c_{ij}^XR_{jn}~.
    \end{split}
\end{gather}
$R_{jn}$ and $L_{nj}$ are simply the components of the right and left eigenvectors of $\vb*{T}_n^{\mathrm{P}}$, respectively, and $b_{ji}^{X} = \vb*{w}_j^T\vb*{b}_i^X$ and $c_{ij}^X = \vb*{c}_i^X\vb*{v}_{j}$ are products of starting vectors and biorthonormalized Lanczos vectors.

\subsection{Generation of a reduced basis}\label{sec:reducedbasis}
The iterative process that is used to calculate sets of excited EOM-CC states is given below. We take the set $J_0$ to contain the indices of already calculated EOM-CC states, and start from the ground state, giving $J_0 = \{0\}$. 
The iterative procedure for the $c$-th EOM-CC state calculation is as follows.
\begin{enumerate}
    \item Choose a chain length, $n_c$, and a subset of the state indices from previous calculations, $I_c \subseteq J_{c-1}$. Also choose a set of operators $X_c$ based on the final states accessed by the operators (see \cref{sec:results}).
    \item Sequences of right and left start vectors, $(\vb*{b}_i^X)_{i \in I_c, X \in X_c}$ and $(\vb*{c}_i^X)_{i \in I_c, X \in X_c}$, are constructed in accordance with \crefrange{eq:b0}{eq:cm}.
    \item The band Lanczos algorithm described in \cref{sec:lanczos} is run with the chain length $k_c$, the field-free Jacobian matrix $\vb*{A}^{(0)}$, and the sequences of starting vectors, constructing the matrix $\vb*{T}_{n_c}^\mathrm{P}$.
    \item After the band Lanczos algorithm has finished, the eigenvalues and corresponding right and left eigenvectors of $\vb*{T}_{n_c}^\mathrm{P}$ are calculated. Together, these determine a set of approximate EOM-CC states indexed by $\{m\}$.
    \item States $m$ with eigenvalues $\omega_m^c > \omega^{\text{max}}$ are discarded.
    \item Matrix elements for all operators $X \in X_c$ and combinations of final $m$ and initial states $i \in I_c$ are calculated in accordance with \cref{eq:intransition} and \cref{eq:nitransition}. States $m$ with $X_{im}X_{mi} = S_{im}^X < S^\text{min}$ for all operators and initial states are discarded.
    \item Remaining right and left eigenvectors $m$ of $\vb*{T}_{n_c}^\mathrm{P}$ are transformed to approximate right and left eigenvectors $\vb*{R}_m$ and $\vb*{L}_m$ of $\vb*{A}^{(0)}$ by premultipication by $\vb*{V}_n$ and $\vb*{W}_n^T\vb*{\Delta}_n^{-1}$, respectively. States with vectors linearly independent of previously calculated vectors are stored. The indices of the stored states are added to the previous index set $J_c = J_{c-1} \cup \{m\}$.
\end{enumerate}
The iterative procedure is repeated if states of higher excitation levels are desired. Afterwards, the excited-state Jacobian and overlap matrices, with elements $\widetilde{A}_{mn}^{(0)} = \vb*{L}_m^T\vb*{A}^{(0)}\vb*{R}_n$ and $\widetilde{S}_{mn} = \vb*{L}_m^T\vb*{R}_n$, are constructed in the reduced basis of the approximate eigenvectors. The right and left generalized eigenvalue problems are solved, giving new sets of right and left eigenvectors of $\widetilde{\vb*{A}}^{(0)}$. The dipole and field-free Hamiltonian matrices are then calculated in the basis of both the ground and the newly generated excited states, in accordance with \cref{eq:matrixelements}, and used in
solving the time-dependent problems defined by \cref{eq:righttdeomcc} and \cref{eq:lefttdeomcc}.

\section{Computational details}\label{sec:computational}
Experimental geometries from the NIST database~\cite{Nist} are used for LiH, LiF and C$_2$H$_4$. An optimized geometry from the same database is used for glycine, obtained with the MP2 method with all electrons correlated and the cc-pVTZ basis set. The linear molecules LiH and LiF are aligned along the $z$ axis, as done in Ref.~\cite{PhysRevA.102.023115}. The ethylene molecule is  placed in the $xy$ plane, with the C-C bond along the $x$ axis. The glycine molecule is of $\mathrm{C}_\mathrm{s}$ symmetry for the chosen geometry, with the $xy$ plane as the mirror plane.

In all following calculations, the aug-cc-pCVDZ basis set~\cite{Kendall1992,Woon1995} is used for atoms targeted by the core-exciting pulses; the aug-cc-pVDZ~\cite{Kendall1992} basis set is adopted for the remaining atoms in the molecules. Valence and core states are obtained with the asymmetric band Lanczos algorithm with varying chain lengths as specified in \cref{sec:results}.

Lanczos vectors with Euclidian norms of less than \num{1e-9} are deflated, but this did not occur in any of the calculations. Final excited states that do not have a minimum transition strength of at least \num{1e-7} to any initial state, for any of the operators used to construct the starting vectors, are discarded. This is done to only keep states that give a non-negligible contribution to the dynamics. Also, states with excitation energies above $\omega^\text{max} = \sum_i^{n_\gamma} E_{\gamma_i}^\text{max}$, are discarded, where $E_{\gamma_i}^\text{max} = \omega_i + 8\sigma_i^\omega$ is an estimate of the maximum energy of photon $i$ involved in the $n_\gamma$-photon transition to the desired excited states. The carrier frequency $\omega_i$ and the frequency RMS width $\sigma_i^\omega = 1/(2\sigma_i^t)$ are parameters of the pulse providing photon $i$, see \cref{sec:system}.

Core-valence separation (CVS)~\cite{Cederbaum1980,CVS:2015,Bruno-cvs} projectors were used to calculate core states. A ``core-only'' CVS projector is applied to remove excitations that originate exclusively from valence orbitals. This is done by zeroing out all right and left vector elements that only involve molecular orbitals with energies greater than the energy of the lowest core molecular orbital of a given atom. This yields a Lanczos spectrum starting at the lowest core excitation energy of the chosen edge. A complementary ``valence-only'' CVS projector is used to obtain valence excited states that are orthogonal to the core excited states.

Except for the spectra presented in \cref{fig:lif_unconv}, only sufficiently converged valence and core band Lanczos vectors are used for calculating stationary states and corresponding Hamiltonian and transition moment matrices. This is done by discarding states with either right or left residual norms greater than \num{1e-2} for valence states and \num{1e-1} for core states.

In all calculations, valence states are calculated first, with starting vectors based on the ground state. All accepted valence states are, together with the ground state, used to construct starting vectors for the core state calculations, see \cref{sec:reducedbasis}.

A fixed pump-probe delay of \SI{40}{\au} (about \SI{0.967552}{\femto\second}) is used for both lithium fluoride, lithium hydride and ethylene. The delay is varied for glycine, in order to calculate the transient absorption of the molecule. In all calculations the central time of the probe pulse is set to~0 and the negative central times of the pump pulses are set accordingly.

Unless otherwise stated, integration of the TD-EOM-CC and TDCC equations is done using a Dormand-Prince 5(4) integration scheme~\cite{DORMAND198019} with a maximum time step of \SI{0.1}{\au}, and maximum and minimum local errors of \num{1e-7} and \num{1e-9}, respectively, see \cref{sec:integration}. Each component of the time-dependent dipole moment expectation value and electric field vectors are multiplied with the Hann window before Fourier transformation.

All calculations are performed using a development version of the $e^T$ program~\cite{et-release}.

\section{Results and discussion} \label{sec:results}
\subsection{Lithium fluoride: convergence and non-linear pump interaction}
\label{sec:lif}
\begin{figure*}
    \centering
    \includegraphics[width=6.75in]{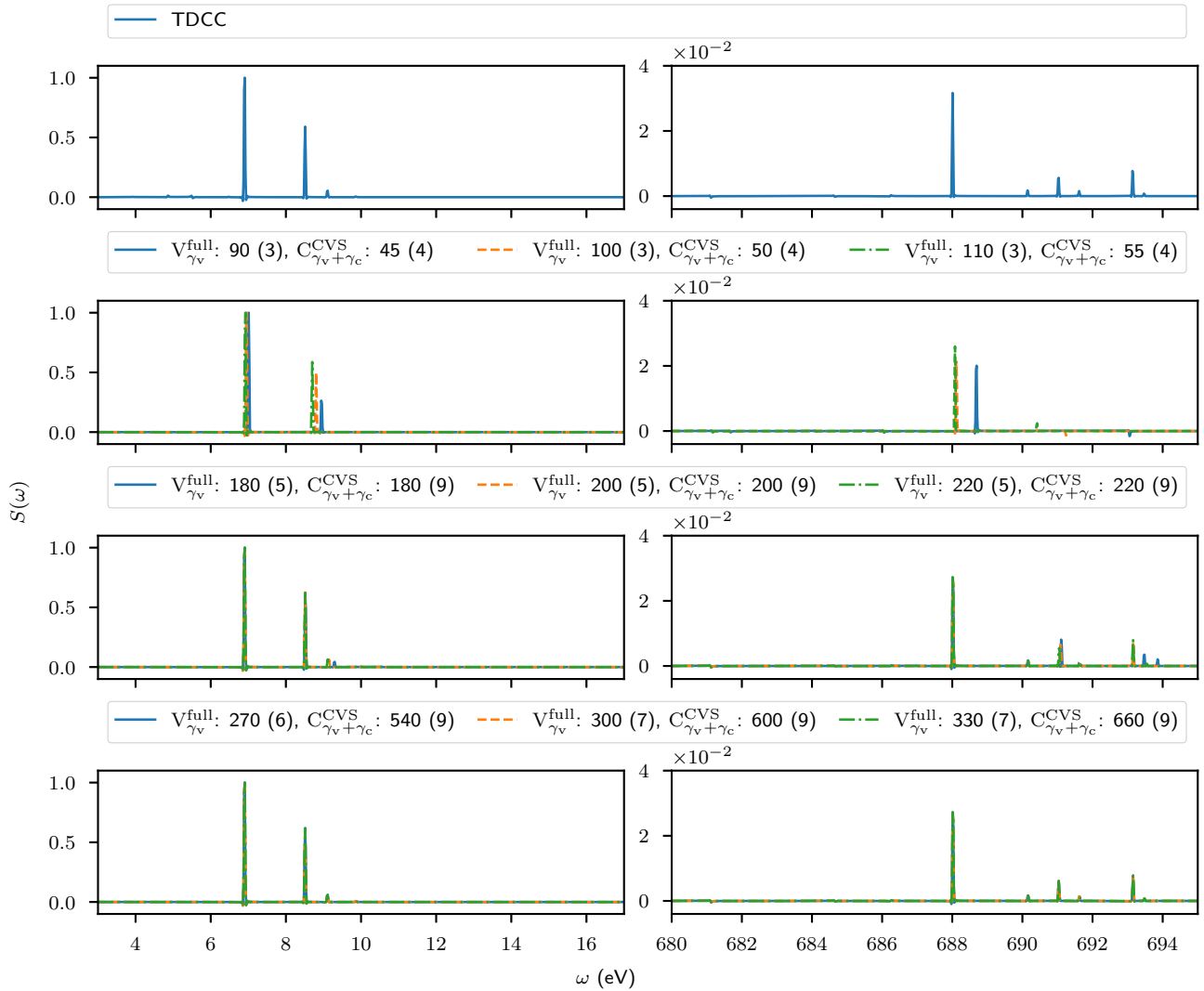}
    \caption{LiF pump-probe absorption $S(\omega)$ as a function of frequency $\omega$ in the valence and core regions, normalized by the tallest peaks in the spectra. The TDCC results are shown in the top left and right panels. TD-EOM-CC results, calculated at different band Lanczos chain lengths, are shown in the lower panels. EOM-CC valence (V) states are calculated in the full projection space, while the core (C) states are calculated within the CVS approximation. Valence states energetically inaccessible by a single pump photon $\gamma_\mathrm{v}$ are discarded, and so are core states energetically inaccessible by subsequent absorption of a probe photon $\gamma_\mathrm{v}+\gamma_\mathrm{c}$. The chain lengths of the calculations are given, together with the number of converged states (in brackets).}
    \label{fig:lif_unconv}
\end{figure*}
\begin{figure*}
    \centering
    \includegraphics[width=6.75in]{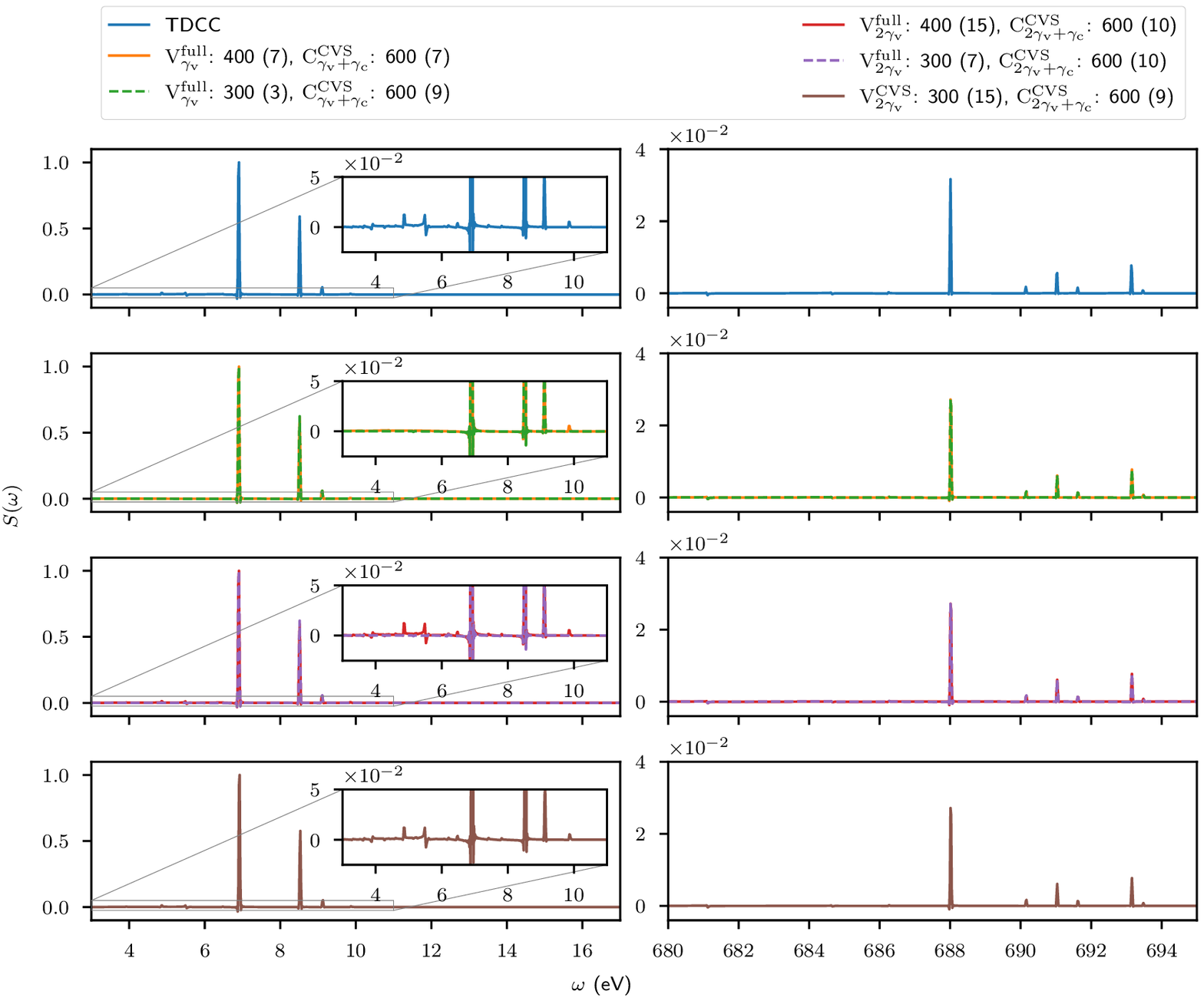}
    \caption{LiF pump-probe absorption $S(\omega)$ as a function of frequency $\omega$ in the valence and core regions, normalized by the tallest peaks in the spectra. The TDCC results are shown in the top left and right panels. TD-EOM-CC results, calculated at different band Lanczos chain lengths, are shown in the lower panels. EOM-CC valence (V) states are calculated in the full projection space (middle panels) or within the CVS approximation (bottom panels), while the core (C) states are calculated within the CVS approximation only. Valence states energetically inaccessible by a single or two pump photons, $\gamma_\mathrm{v}$ or $2\gamma_\mathrm{v}$, are discarded, and so are core states energetically inaccessible by subsequent absorption of a probe photon, $\gamma_\mathrm{v}+\gamma_\mathrm{c}$ or $2\gamma_\mathrm{v}+\gamma_\mathrm{c}$. The chain lengths of the calculations are given, together with the number of converged states (in brackets).}
    \label{fig:lif}
\end{figure*}
When discussing the applicability of the band Lanczos algorithm for modeling attosecond pump-probe processes, a key question is how spectra are affected by the chain length used.
With this in mind, a single TDCC LiF pump-probe absorption spectrum, calculated with the RK4 integrator and fixed time steps of \SI{0.005}{\au}, is in \cref{fig:lif_unconv} compared to TD-EOM-CC spectra calculated with the Dormand-Prince 5(4) integration scheme and various band Lanczos chain lengths. In all calculations, the pulses have the parameters used for the LiF spectra in Sec. III B of~\cite{PhysRevA.102.023115}, where the F \textit{K}-edge is targeted by the probe pulse. All states with energies inaccessible by the absorption of one photon from each pulse are discarded from the TD-EOM-CC calculations.

For lower chain lengths, the peaks of the band Lanczos spectra shown in \cref{fig:lif_unconv} both shift and scale significantly with variations in the chain length, indicating that excitation energies and dipole matrix elements are badly converged. The convergence generally improves with the chain length, and low-energy high-amplitude peaks seem to converge first. Higher chain lengths are needed for good convergence of high-energy low-amplitude peaks, as expected from the convergence behavior of Lanczos algorithms.

As demonstrated, the inclusion of badly converged states can give spectral peaks with incorrect positions and amplitudes. In addition, these states can also increase the cost of matrix element calculation and propagation, decrease the convergence rate of consecutive band Lanczos calculations, and cause serious numerical instabilities during propagation. In order to avoid these adverse effects, states with badly converged right or left vector residual norms will be discarded in the following band Lanczos calculations.

In \cref{fig:lif}, the aforementioned TDCC LiF pump-probe absorption spectrum is compared to TD-EOM-CC spectra from converged states only. Note that the three most dominant peaks in the TDCC spectrum are present in the green spectrum, which is calculated with a valence chain length of 300, but a chain length of \num{400} is needed in order to converge the short peak at around \SI{10}{\electronvolt}. The low amplitude peaks below and around the tall peak at around \SI{6.9}{\electronvolt} are missing.

In an earlier work~\cite{PhysRevA.102.023115}, we speculated that the smaller peaks below \SI{6.9}{\electronvolt} in the pump-only LiF spectrum could originate from two-photon absorption. This claim was later discussed by Pedersen et al.~\cite{doi:10.1021/acs.jctc.0c00977}, where the TDCC state of LiF interacting with the pump pulse was analyzed in terms of stationary state populations. Their analysis supports the interpretation that two photons are absorbed from the pump pulse.

In order to take two-photon absorption into account, spectra are recalculated with the inclusion of valence states energetically accessible by two pump photons and core states accessible by an additional probe photon. The corresponding results obtained with chain lengths of 300 and 400 are shown in purple and red in \cref{fig:lif}, respectively. Note that the 300 valence chain length spectrum still lacks the smaller features of the TDCC spectrum, but the 400 valence chain length spectrum is practically indistinguishable from the TDCC one. This similarity corroborates the claim that two photons are absorbed from the pump pulse. Furthermore, the results demonstrate that reduced-basis TD-EOM-CC can faithfully reproduce TDCC results in particular systems, even when non-linear interactions are involved. The embedded Dormand-Prince 5(4) integrator is seen to perform well for TD-EOM-CC.

The bottom panel of \cref{fig:lif} demonstrates the use of the valence-only CVS projector to calculate the valence states. The approximation seems to improve the rate of convergence with respect to chain length, as a length of \num{300} is enough to retrieve all the features of the TDCC spectrum while a higher number is necessary in the non-projected case. This improved convergence can be explained by the reduction in dimension from projecting out transitions from core orbitals. Moreover, since the approximation does not seem to lead to significant scaling or shifting of the valence peaks, it is  adopted in the following calculations.

\subsection{Lithium hydride: applicability of the CVS projectors}\label{sec:lih}
\begin{figure*}
    \centering
    \includegraphics[width=6.75in]{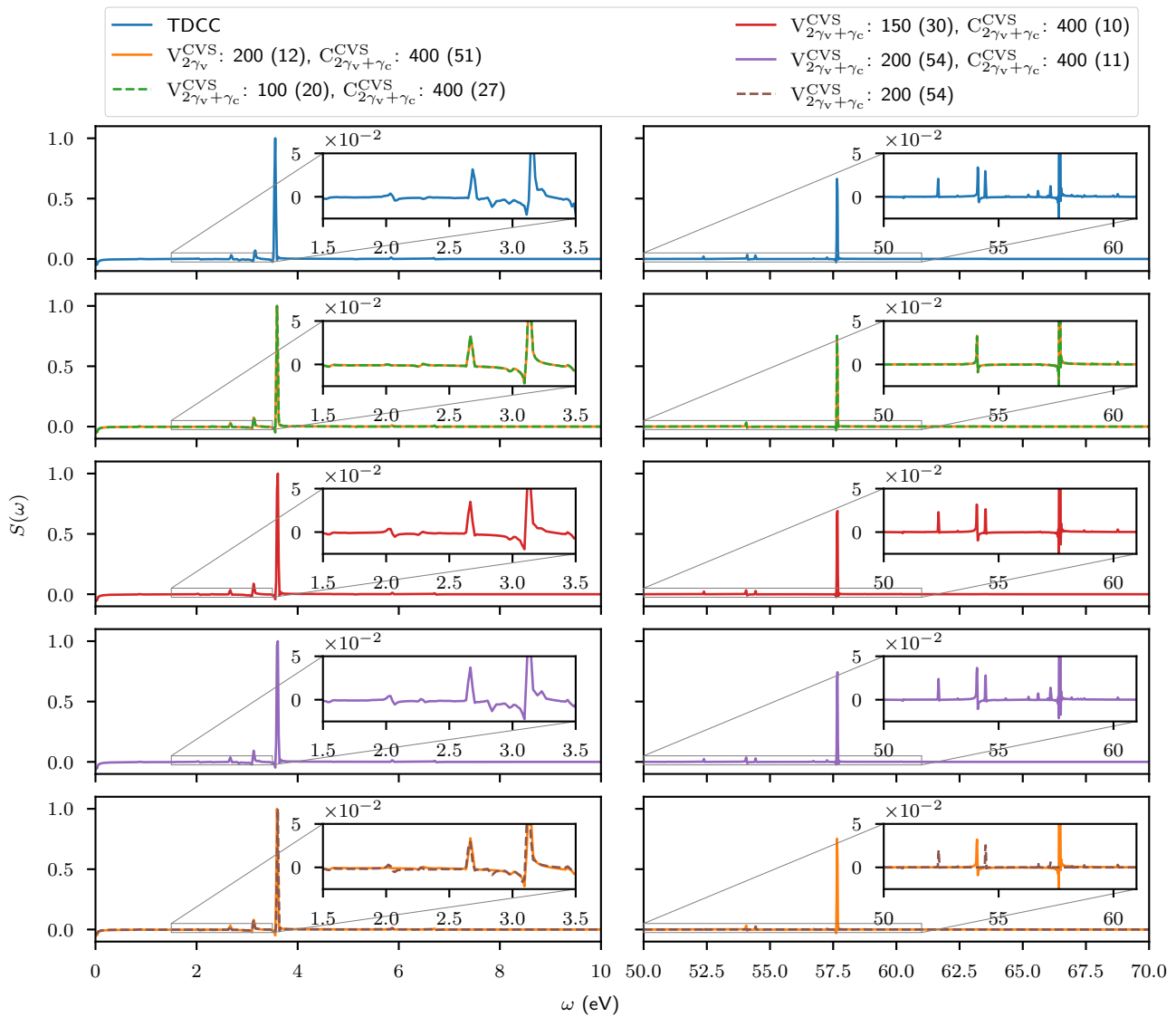}
    \caption{LiH pump-probe absorption $S(\omega)$ as a function of frequency $\omega$ in the valence and core regions, normalized by the tallest peaks in the spectra. The TDCC results are shown in the top left and right panels. TD-EOM-CC results, calculated at different band Lanczos chain lengths, are shown in the lower panels. All EOM-CC valence (V) and core (C) states are calculated within the CVS approximation. For the CVS valence calculation for the results shown in orange, states energetically inaccessible by two pump photons, $2\gamma_\mathrm{v}$ are discarded. For all other band Lanczos calculations, only the states inaccessible by two pump photons and one probe photon $2\gamma_\mathrm{v}+\gamma_\mathrm{c}$ are discarded. The results shown in brown are calculated from CVS valence states only. The chain lengths of the calculations are given, together with the number of converged states (in brackets).}
    \label{fig:lih_rmcore}
\end{figure*}

To further assess the performance of the proposed procedure, as well as the applicability of the CVS projectors, we use the TD-EOM-CC procedure to model the interaction of the lithium hydride molecule with the pump-probe pair described in Sec. III A of Ref.~\cite{PhysRevA.102.023115}. Li \textit{K}-edge spectra are notoriously difficult to describe accurately due to the small energy separation between the valence and core excitation regions. This can be considered a challenging test case for the applicability of the core-valence separation scheme.

A comparison between TD-EOM-CC and TDCC spectra is given in \cref{fig:lih_rmcore}, where the latter is taken from Ref.~\cite{PhysRevA.102.023115}. In all core state calculations a fixed band Lanczos chain length of \num{400} is used. However, the number of converged core states, given in brackets, differs due to the different starting vectors employed.

Since TD-EOM-CC with energy-limited valence and core states successfully reproduced the TDCC spectrum of LiF in \cref{sec:lif}, a similar procedure is attempted for calculating the TD-EOM-CC LiH spectrum. That is, valence states inaccessible by two pump photons and core states inaccessible by an additional probe photon are discarded. The results are shown in orange in the second topmost panels of \cref{fig:lih_rmcore}. The spectrum lacks some of the weaker features in the valence excitation energy region, and, more notably, many of the dominant features in the core excitation region. In other words, a characteristic of the LiH molecule seemingly prevents us from reproducing the TDCC spectrum using the procedure in the previous section. In the following, we argue that the Li \textit{K}-edge in LiH involves states that cannot be obtained with the core-only CVS projector alone, as they do not correspond to core excitations.

TD-EOM-CC spectra calculated with states obtained with the valence-only CVS projector, energetically accessible by two pump and one probe photons, are shown the three middle rows of panels in \cref{fig:lih_rmcore}. The valence chain lengths used are \num{100} (green), \num{150} (red) and \num{200} (purple). The number of converged states in the valence region increases with the chain length. Remarkably, increasing the valence chain length also leads to additional peaks in the core region, illustrated in the right panels. This demonstrates that, apart from the two intense peaks obtained at about \SI{54.1}{\electronvolt} and \SI{57.7}{\electronvolt}, the other peaks are of pure valence excitation character.

The necessity to include high-energy states calculated with the valence-only CVS projector is further validated by superimposing a spectrum exclusively from valence-only CVS states (brown) with the spectrum calculated from energy-limited valence and core states (orange), shown in the bottom panels of \cref{fig:lih_rmcore}. The peaks of the composite spectrum are in good agreement with the TDCC ones. Therefore, we assert that use of both the core-only and the complementary valence-only CVS projectors is necessary in order to accurately capture the spectral features around the Li \textit{K}-edge in LiH. Note that this should not be taken as a failure of the CVS projectors \textit{per se} but as a consequence of 
the peculiar electronic structure of LiH. In fact, the high-energy states of pure valence character can be more easily calculated in the dimension reduced by the valence-only CVS projector.

\subsection{Ethylene: non-linear pump interaction for a different symmetry group}\label{sec:c2h4}
\begin{figure*}
    \centering
    \includegraphics[width=6.75in]{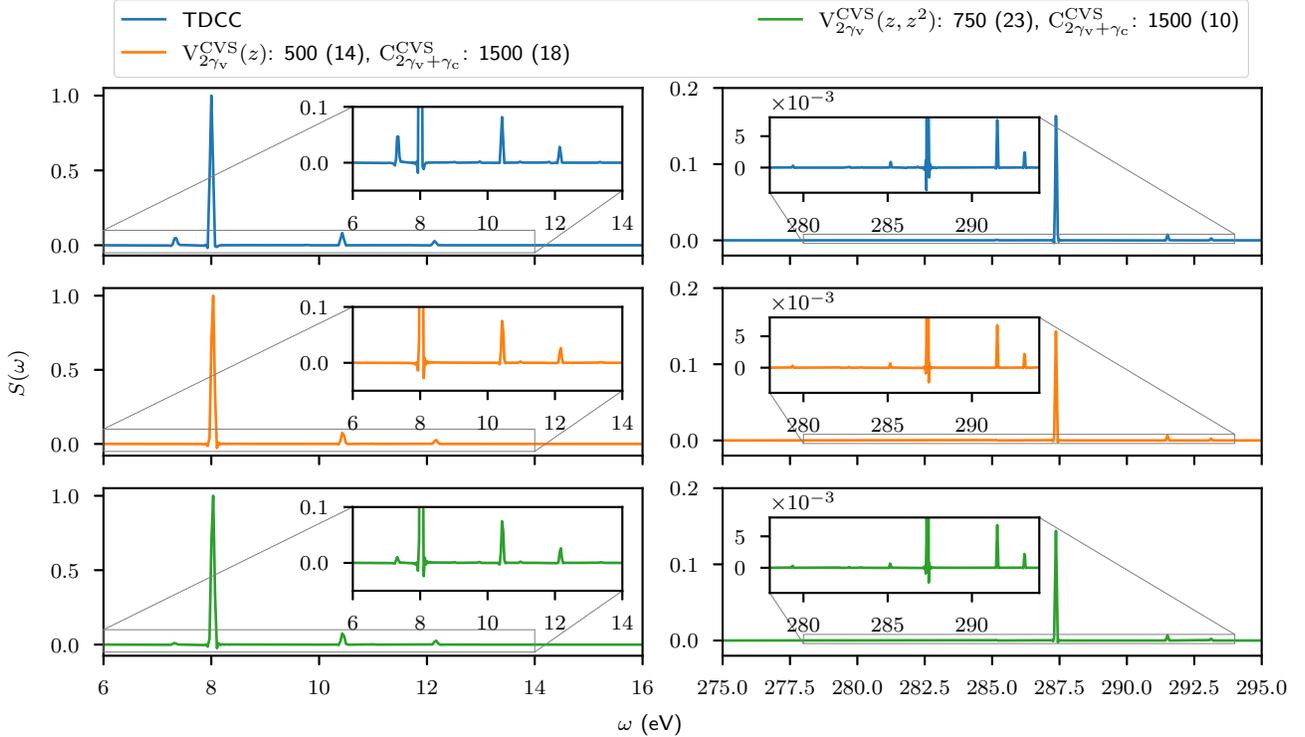}
    \caption{Ethylene pump-probe absorption $S(\omega)$ as a function of frequency $\omega$ in the valence and core regions, normalized by the tallest peaks in the spectra. The TDCC results are shown in the top left and right panels, and TD-EOM-CC results in the lower panels. In the middle panels, a ground state $z$ dipole operator starting vector is used for calculating the valence (V) states. In the bottom panels, both ground-state $z$ dipole operator and $z^2$ quadrupole operator starting vectors are used. All valence and core (C) states are calculated within the CVS approximation. The chain lengths of the calculations are given, together with the number of converged states (in brackets).}
    \label{fig:ethylene}
\end{figure*}

For ethylene, RMS widths of the  $z$-polarized pump and $x$-polarized probe pulses are set to \SI{10}{\au} and \SI{5}{\au}, corresponding to intensity FWHM durations of about \SI{403}{\atto\second} and \SI{201}{\atto\second}, respectively. The carrier frequency of the pump pulse is set to \SI{8.003337}{\electronvolt}, and the probe pulse to \SI{285.608346}{\electronvolt} (C \textit{K}-edge). The time-dependent state is propagated with the Dormand Prince 5(4) integration scheme, from \SIrange{-2500}{2500}{\au} of time. The TDCC spectrum, shown in the top panel \cref{fig:ethylene}, is characterised by four dominant peaks in the valence excitation region. A low amplitude peak at around \SI{7.3}{\electronvolt} is present in the TDCC spectrum, but missing in the TD-EOM-CC spectrum calculated with a $z$ dipole operator starting vector, shown in the middle panels. In accordance with the interpretation of the spectrum of LiF in \cref{sec:lih}, we attribute the missing peak to a two-photon excitation process, even though valence states energetically accessible by two photons are included. Note that quadratic functions of the $z$ dipole operator belong to the $\mathrm{A}_g$ representation of $\mathrm{D}_{2\mathrm{h}}$, the point group of ethylene for the chosen geometry. Hence, we should not expect the single starting vector, belonging to the $\mathrm{B}_{1\mathrm{u}}$ representation, to facilitate the convergence of the two-photon peaks.

In order to mimic the two pump photon absorption process, we include a starting vector constructed from the $z^2$ quadrupole operator in the valence state calculation. The results, shown in the bottom panels of \cref{fig:ethylene}, now capture the two-photon peak at around \SI{7.3}{\electronvolt}.  The amplitude yielded by TD-EOM-CC is, however, underestimated compared to the TDCC one, which can indicate that more secondary valence excited states should be included in the computation. It may also be a signature of fundamental differences in two-photon absorption as described by TD-EOM-CC and TDCC.
\subsection{Glycine: transient absorption}
\label{sec:glycine}
\begin{figure*}
    \centering
    \includegraphics[width=6.75in]{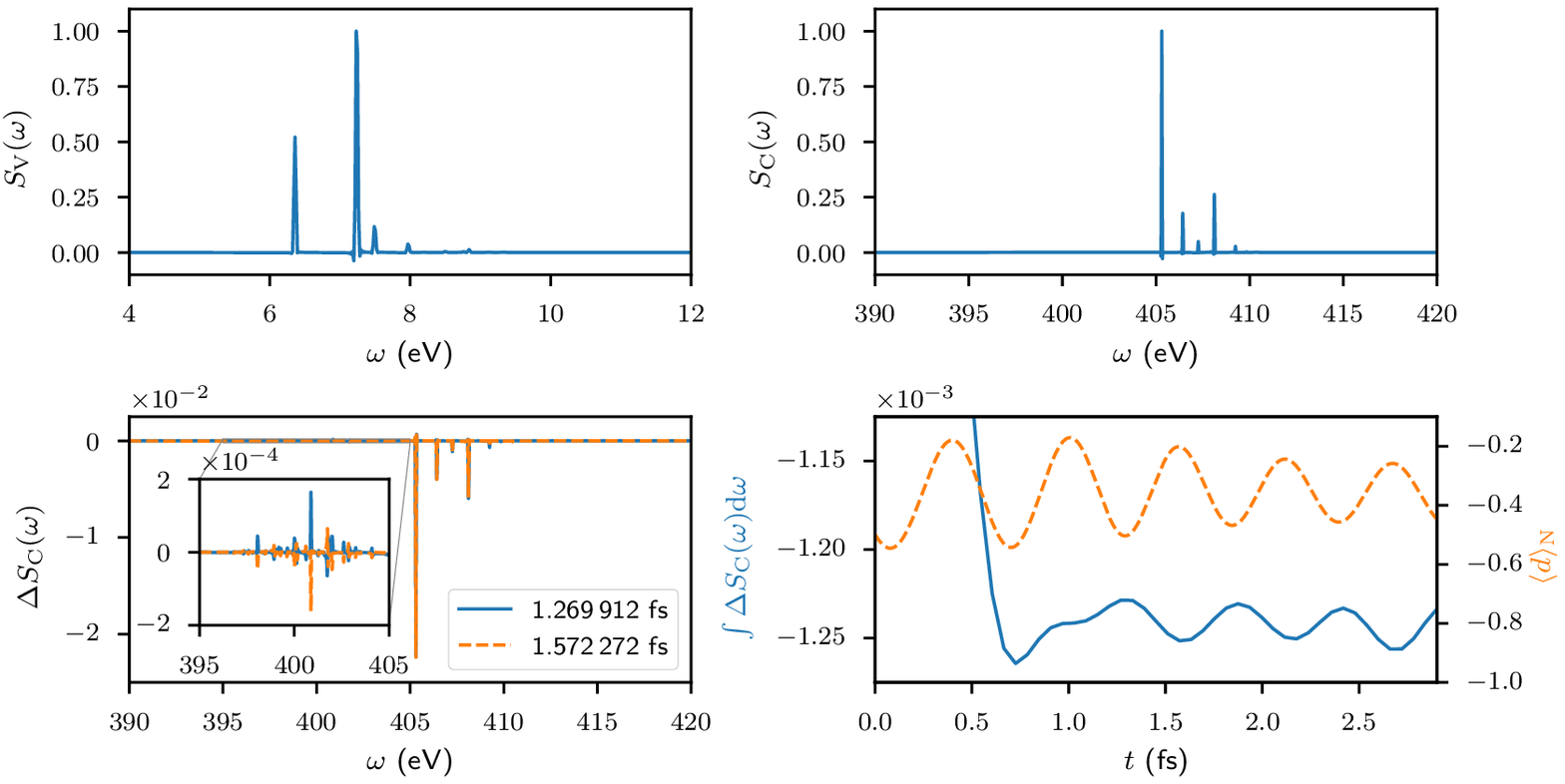}
    \caption{Glycine. Top left: pump pulse absorption from the ground state $S_\mathrm{V}(\omega)$, normalized by the tallest peak in the spectrum. Top right: probe pulse absorption from the ground state $S_\mathrm{C}(\omega)$, normalized by the tallest peak in the spectrum. Bottom left: pump-probe absorption minus the probe absorption from the ground state $\Delta S_\mathrm{C}(\omega) = S(\omega) - S_\mathrm{C}(\omega)$, normalized by the tallest peak in the $S_\mathrm{C}(\omega)$ spectrum. The results for two different pump-probe delays are shown. Bottom right: in blue, the numerically integrated probe absorption difference $\int \Delta S_\mathrm{C}(\omega)\dd{\omega}$ is shown as a function of pump-probe delay. In orange, the dipole induced by the pump pulse in the direction from the center of mass to the N atom, $\ev{d}_\mathrm{N}$, is shown as a function of time after the center of the pump pulse.}
    \label{fig:glycine}
\end{figure*}
As a final example, we use computational procedure to model attosecond transient absorption by the glycine molecule. The RMS widths of $(0.490072x + 0.871682y)$-polarized pump and $z$-polarized probe pulses are set to \SI{20}{\au} and \SI{10}{\au}, corresponding to intensity FWHM durations of about \SI{806}{\atto\second} and \SI{403}{\atto\second}, respectively. The carrier frequency of the pump pulse is set to \SI{6.358895}{\electronvolt}, and the probe pulse to \SI{405.305168}{\electronvolt} (N \textit{K}-edge). The time-dependent state is propagated with the Dormand Prince 5(4) integration scheme, from \SIrange{-5000}{5000}{\au} of time.

A single band Lanczos calculation is used for constructing the valence states, where states energetically inaccessible by two-photon transitions are discarded. Note that we do not need to use quadrupole operators in order to get two-photon valence states of glycine in the reduced basis, as otherwise done for the ethylene valence states, since both linear and quadratic functions of $x$ and $y$ dipole operators belong to the $\mathrm{A}'$ representation of $\mathrm{C}_\mathrm{s}$. A valence state calculation, with ground-state start vectors and a chain length of \num{1500}, gives \num{17} converged states. A subsequent core state calculation, with ground and valence state start vectors and a chain length of \num{3000}, gives \num{20} converged states.

As a note of caution, the maximum energy of the valence states is \SI{10.458724}{\electronvolt}, which is below the double frequency of the carrier photons. This indicates that two-photon absorption is not properly accounted for by the reduced basis, meaning that smaller features may be missing in the spectra.

In the top left panel of \cref{fig:glycine}, the absorption of the pump pulse is shown as a function of frequency. Even though the glycine molecule is substantially larger than the other molecules considered, the spectrum is still dominated by a small number of peaks. The number of dominant peaks is also smaller than the number of converged states in the basis (\num{17}). The spectrum of the absorption of the probe pulse by the ground state, shown in the top right panel, also has fewer dominant peaks than the number of converged states (\num{20}).

In order to calculate the transient absorption of the probe pulse by the glycine molecule, absorption spectra are calculated with the pump-probe setup used for the other molecules, with pump-probe delays varying from \SI{0}{\femto\second} to \SI{2.902656}{\femto\second} in intervals of \SI{60.472}{\atto\second}. The reduced basis energies and dipole matrix elements do not have to be recalculated between the different TD-EOM-CC calculations, since these are independent of the pump-probe delay. Difference spectra are then calculated by subtracting the ground-state probe absorption spectrum from the pump-probe absorption spectra, before normalizing by the tallest peak in the ground-state probe spectrum. In the bottom left panel, the difference spectra at \SI{1.269912}{\femto\second} and \SI{1.572272}{\femto\second} are shown in blue and orange, respectively. Both spectra are dominated by negative peaks, indicative of ground state bleaching. In addition, the spectra vary slightly with pump-probe delay, which is particularly visible for the peaks that are not energetically accessible from the ground state, e.g. in the energy range from \SIrange{395}{405}{\electronvolt} shown in the inset.

In order to quantify the delay-dependent difference in the absorption of the full probe pulse, each of the \num{49} pump-probe difference spectra are numerically integrated from \SIrange{390}{420}{\electronvolt} using the trapezoidal rule. The results are shown as function of pump-probe delay in the bottom right panel of \cref{fig:glycine}, in blue. Note that the absorption difference is smaller for short pump-probe delays, which can be explained by the fact that ground state bleaching happens gradually during the pump pulse interaction. A shorter pump-probe delay implies that the molecule is probed while bleaching still occurs, which can lead to a smaller difference between the pump-probe absorption and the ground state probe absorption.

We have also calculated the pump-induced time-dependent dipole moment in the direction from the center-of-mass to the center of the N atom, as a way of quantifying the migration of charge between the end containing the N atom and the opposite end of the molecule. The dipole moment is shown in the bottom right panel of \cref{fig:glycine}, in orange.

Note that the dominant periods of both the time-dependent dipole moment and the integrated absorption (after \SI{1}{\femto\second}), shown in the bottom right panel of \cref{fig:glycine}, fall within \SI{0.57\pm0.04}{\femto\second}. This indicates that the pump-induced TD-EOM-CC state is a coherent superposition dominated by states with energy differences of \SI{7.3\pm0.6}{\electronvolt}, which is in agreement with the ground state pump absorption spectrum (top left). It also indicates that the dominant features of the time-dependent charge migration and the delay-dependent \textit{K}-edge absorption are correlated and can be measured with phase-controlled pulses with finite duration, as has previously been demonstrated for instantaneous pulses~\cite{PhysRevA.88.013419,PhysRevA.90.023414}.

\section{Conclusion} \label{sec:conclusion}
We have demonstrated the use of the asymmetric band Lanczos algorithm to generate reduced TD-EOM-CC bases for various molecules, taking the characteristics of pulses suitable for probing attosecond phenomena into account. The specific start vectors used in the calculations direct the band Lanczos algorithm towards states that are useful for representing the interactions. The start vectors also allow for the affordable calculation of transition strengths, which are used, together with excitation energies, to automatically select the reduced basis. The basis is further reduced by removing unconverged states.

In \cref{sec:lif}, we demonstrated how lithium fluoride spectral peaks can converge towards peaks calculated with TDCC by increasing the band Lanczos chain length and taking a sufficient number of relevant states into account. In particular, we showed that two-photon absorption has to be taken into account in order to reproduce the smaller features of the TDCC spectrum, as speculated in Ref.~\cite{PhysRevA.102.023115}.

In \cref{sec:lih}, we demonstrated that the core-only CVS projector eliminates several of the peaks around the \textit{K}-edge of lithium in lithium hydride. The missing peaks can be captured with the complementary valence-only CVS projector, which enabled us to target high-energy states of pure valence character. This observation indicates that care should be taken when the CVS scheme is used for light elements such as lithium, where the energy separation of the core and valence orbitals is small, 
so that pure valence excitations can fall within the region of core excitations.

In \cref{sec:c2h4}, we used starting vectors constructed from both dipole and quadrupole operators, in order to converge ethylene valence states that are dark with respect to one-photon transitions from the ground state.

In \cref{sec:glycine}, pump and probe pulses with varying time delays were used to assess the transient absorption of a \textit{K}-edge probe pulse as a function of pump-probe delay. We showed how the transient absorption seems to correlate with the migration of charge induced by the pump, and how both quantities seem to reveal the dominant timescale in the coherent superposition.

\begin{acknowledgments}
We acknowledge Sarai D. Folkestad for her role in implementing the simple Lanczos algorithm and density-based excited-state transition moments, available in the $e^T$ program, which serve as precursors to the implementations here. We acknowledge financial support from The Research Council of Norway through FRINATEK Project Nos. 263110 and 275506, from the Marie Sk{\l}odowska-Curie European Training Network \textit{COmputational Spectroscopy In Natural sciences and Engineering (COSINE)}, Grant Agreement No. 765739, and from the Independent Research Fund Denmark, DFF-FNU Research Project 2 No. 7014-00258B. Computing resources through UNINETT Sigma2---the National Infrastructure for High Performance Computing and Data Storage in Norway (Project No. NN2962k) and through the Center for High Performance Computing (CHPC) at SNS are also acknowledged. Finally, the COST Action CA18222, \textit{AttoChem}, is acknowledged.
\end{acknowledgments}

\section*{Author's contributions}
S.C. and H.K. conceptualized and supervised the project. T.M., S.C. and A.S.S. implemented and optimized the band Lanczos solver. A.B. and A.S.S. did the initial TDCC implementation, and A.S.S. the TD-EOM-CC extension. A.C.P. implemented procedures for calculating eigenstates and matrix elements of combined sets of EOM-CC states. A.S.S. and T.M. carried out the calculations and interpreted the results. All authors contributed to the writing of the manuscript.

\appendix
\section{Rewriting EOM-CC matrix element expressions}\label{sec:matrixelements}
The matrix element $X_{ij}$ of the operator $X$ and the left and right vectors of EOM-CC states $i$ and $j$, respectively, can be written as
\begin{equation}
    \label{eq:rewrittenmatrixelements}
    \begin{split}
        X_{ij}
        &= \bra*{\widetilde{\psi}_i}X\ket{\psi_j} \\
        &= \sum_{\kappa\lambda}l_{i\kappa}\bra{\kappa}\bar{X}\ket{\lambda}r_{\lambda j} \\
        &= \sum_\nu l_{i0}\bra{\HF}\bar{X}\ket{\nu}r_{\nu j} + \sum_{\mu\nu}l_{i\mu}\bra{\mu}\bar{X}\ket{\nu}r_{\nu j} \\
        &\phantom{{}={}}+ \bigg(l_{i0}X_{00} + \sum_\mu l_{i\mu}\xi_\mu^X\bigg)r_{0 j} \\
        &= \sum_\nu l_{i0}\bra{\HF}\comm{\bar{X}}{\tau_\nu}\ket{\HF}r_{\nu j} \\
        &\phantom{{}={}} + \sum_{\mu\nu}l_{i\mu}\Big({}^{\LR}A_{\mu\nu}^X+ \bra{\mu} \tau_\nu\bar{X}\ket{\HF}\Big)r_{\nu j} \\
        &\phantom{{}={}}+ \bigg(l_{i0}X_{00} + \sum_\mu l_{i\mu}\xi_\mu^X\bigg)r_{0 j}~,
    \end{split}
\end{equation}
where
\begin{gather}
    {}^{\LR}A_{\mu\nu}^X = \bra{\mu}\comm{\bar{X}}{\tau_\nu}\ket{\HF}~, \\
    X_{00} = \bra{\HF}\bar{X}\ket{\HF}~, \\
    \xi_\mu^X = \bra{\mu}\bar{X}\ket{\HF}~.
\end{gather}
\subsection{Left excited and right ground state}
The left excited state $m$ has the reference component $l_{m0} = 0$ (cf. \cref{eq:leftrefcomponent}), while the right ground state has components $r_{00} = 1$, $r_{\mu0} = 0$ (cf. \cref{eq:rightgscomponents}). Inserting this into \cref{eq:rewrittenmatrixelements}, we obtain
\begin{equation}
    \label{eq:rightgroundelements}
    X_{m0} = \sum_\mu l_{m\mu}\xi_\mu^X~,
\end{equation}
which is also the expression appearing in CC response theory~\cite{Sonia-cc-lanczos}. Note that this expression is linear in the excited determinant components $l_{m\mu}$.

\subsection{Left ground and right excited state}
The left ground state has the components $l_{00} = 1$, $l_{0\mu} = \bar{t}_\mu$ (cf. \cref{eq:leftgscomponents}), while the right excited state $m$ has the reference component $r_{0m} = -\sum_\mu \bar{t}_\mu r_{\mu m}$ (cf. \cref{eq:rightrefcomponent}). Inserting this into \cref{eq:rewrittenmatrixelements}, we obtain
\begin{equation}
    \label{eq:leftgroundelements}
    \begin{split}
        X_{0n}
        &= \sum_\nu \bigg({}^{\LR}\eta_\nu + \sum_\mu \bar{t}_\mu\bra{\mu} \tau_\nu\bar{X}\ket{\HF}\bigg)r_{\nu n} \\
        &\phantom{{}={}} - \bigg(X_{00} + \sum_\mu\bar{t}_{\mu}\xi_\mu^X\bigg)\sum_\nu \bar{t}_\nu r_{\nu n}~,
    \end{split}
\end{equation}
where
\begin{equation}
     {}^{\LR}\eta_\nu = \bra{\HF}\comm{\bar{X}}{\tau_\nu}\ket{\HF} + \sum_\mu\bar{t}_\mu{}^{\LR}A_{\mu\nu}^X~.
\end{equation}
The term $\sum_\nu {}^{\LR}\eta_\nu r_{\nu n}$ appears in CC response theory~\cite{Sonia-cc-lanczos}, and the other terms are specific to EOM-CC. \Cref{eq:leftgroundelements} is equivalent to Eq. (65) in~\cite{doi:10.1063/1.466321} and can also be written as
\begin{equation}
    \label{eq:compare}
    X_{0n} = \sum_\nu \Big({}^{\EOM}\eta_\nu^X - X_{00} \bar{t}_\nu\Big) r_{\nu m}~,
\end{equation}
where
\begin{equation}
    \begin{split}
        {}^{\EOM}\eta_\nu^X
        &= {}^{\LR}\eta_\nu^X + \sum_\mu \bar{t}_{\mu}\bra{\mu}\tau_\nu\bar{X}\ket{\HF} \\
        &\phantom{{}={}}-\bigg(\sum_\mu\bar{t}_\mu\xi_\mu^X\bigg) \bar{t}_\nu
    \end{split}
\end{equation}
cf. Eq. (18) of Ref.~\cite{CPP:RIXS:CC}. Note that \cref{eq:compare} is linear in the excited determinant components $r_{\nu m}$.

\subsection{Left and right excited states}
The left excited state $m$ has the reference component $l_{m0} = 0$, while the right excited state $n$ has $r_{0n} = -\sum_\mu \bar{t}_\mu r_{\mu n}$ (cf. \crefrange{eq:rightrefcomponent}{eq:leftrefcomponent}). Inserting this into \cref{eq:rewrittenmatrixelements}, we obtain
\begin{equation}
\label{eq:excitedelements}
    \begin{split}
        X_{mn}
        &= \sum_{\mu\nu}l_{m\mu}\Big({}^{\LR}A_{\mu\nu}^X + \bra{\mu} \tau_\nu\bar{X}\ket{\HF}\Big)r_{\nu n} \\
        &\phantom{{}={}} - \sum_\mu l_{m\mu}\xi_\mu^X\sum_\nu \bar{t}_\nu r_{\nu n} \\
        &= \sum_{\mu\nu}l_{m\mu}\Big({}^{\LR}A_{\mu\nu}^X + \bra{\mu} \tau_\nu\bar{X}\ket{\HF} - \xi_\mu^X\bar{t}_\nu\Big)r_{\nu n} \\
        &= \sum_{\mu\nu}l_{m\mu}\Big({}^{\EOM}A_{\mu\nu}^X + \delta_{\mu\nu}X_{00} - \xi_\mu^X\bar{t}_\nu\Big)r_{\nu n}~,
    \end{split}
\end{equation}
where
\begin{equation}
    \begin{split}
        {}^{\EOM}A^X_{\mu\nu} &= \bra{\mu}\bar{X}\ket{\nu} - \delta_{\mu\nu}X_{00} \\
        &= {}^{\LR}A_{\mu\nu}^X + \bra{\mu}\tau_\nu \bar{X}\ket{\HF} - \delta_{\mu\nu}X_{00}
    \end{split}
\end{equation}
cf. Eq. (20) in Ref.~\cite{CPP:RIXS:CC}.

The term $\sum_{\mu\nu}l_{m\mu}{}^{\LR}A_{\mu\nu}^Xr_{\nu n}$ appears in CC response theory~\cite{bruno-transient}, and the other terms are specific to EOM-CC. Note that all matrix elements in \cref{eq:excitedelements} are linear in both $l_{m\mu}$ and $r_{\nu n}$.

\section{Integration scheme}\label{sec:integration}
In order to limit the error of the time-dependent results, the integration of the TDCC and TD-EOM-CC equations is done with the embedded Dormand-Prince method of order 5(4)~\cite{DORMAND198019}. This method yields both fourth- and fifth-order accurate solutions at each time step, and is specified by the Butcher tableau~\cite{DORMAND198019,iserles_2008}
\begingroup
\renewcommand*{\arraystretch}{1.5}
\renewcommand*{\arraycolsep}{2pt} 
\begin{align}
    \begin{array}{c|ccccccc}
        0 &&&&&&& \\
        \frac{1}{5} & \frac{1}{5} &&&&&& \\
        \frac{3}{10} & \frac{3}{40} & \frac{9}{40} &&&&& \\
        \frac{4}{5} & \frac{44}{45} & -\frac{56}{15} & \frac{32}{9} &&&& \\
        \frac{8}{9} & \frac{19372}{6561} & -\frac{25360}{2187} & \frac{64448}{6561} & -\frac{212}{729} &&& \\
        1 & \frac{9017}{3168} & -\frac{355}{33} & \frac{46732}{5247} & \frac{49}{176} & -\frac{5103}{18656} && \\
        1 & \frac{35}{384} & 0 & \frac{500}{1113} & \frac{125}{192} & -\frac{2187}{6784} & \frac{11}{84} & \\
        \hline
        & \frac{35}{384} & 0 & \frac{500}{1113} & \frac{125}{192} & -\frac{2187}{6784} & \frac{11}{84} & 0 \\
        & \frac{5179}{57600} & 0 & \frac{7571}{16695} & \frac{393}{640} & -\frac{92097}{339200} & \frac{187}{2100} & \frac{1}{40}
    \end{array}
\end{align}
\endgroup
where the next-to-last and last rows give the coefficients of the fifth and fourth order solutions, respectively. Although the method has seven stages, its \textit{first same as last} property assures that only six function evaluations are needed per time step.

The Euclidean distance between the solutions gives a fourth-order estimate of the local integration error,
\begin{equation}
    \epsilon_{\mathcal{O}(4)} = \norm{\vb*{y}_{\mathcal{O}(5)} - \vb*{y}_{\mathcal{O}(4)}}_2~.
\end{equation}
This local error estimate is kept below a given maximum value by adapting the time step during the integration. The fifth order solution is accepted as the solution at the beginning of the next step whenever the error estimate satisfies this condition.

The following adaptive time stepping scheme was designed, implemented and used together with the Dormand-Prince 5(4) method for the relevant calculations in \cref{sec:results}. At the start of the integration, the step size is set to a given maximum value. During the integration, the variable step size is halved, and the integration step redone, whenever the error estimate exceeds the given maximum error. After a successful integration step, the step size is doubled whenever the error estimate is below a given minimum value, provided that the doubled step size is smaller than the maximum step size and also a submultiple of the elapsed time. This in order to increase the efficiency of the integration while ensuring that the solution is evaluated at times corresponding to integer increments of the maximum time step size. Evaluation of time-dependent observables is done using the solutions at these integer increments.

\bibliographystyle{apsrev4-2}
\bibliography{bibliography}
\end{document}